\documentclass[aps, pra, reprint, superscriptaddress]{revtex4-1}
\usepackage{latexsym}
\usepackage{amsthm}
\usepackage{amssymb}
\usepackage{amsmath}
\usepackage[all]{xy}
\usepackage{float}
\usepackage{graphicx}
\usepackage{subfigure}
\usepackage{dcolumn}
\usepackage{bm}
\usepackage{bbm}
\usepackage[usenames, dvipsnames]{xcolor}
\usepackage{amsfonts}
\usepackage{cases}
\usepackage[a4paper, pagebackref=false, colorlinks=true,
linkcolor=blue, citecolor=blue,
pdfauthor={ },
pdftitle={ },
pdfsubject={ },
pdfkeywords={ }]{hyperref}

\begin{document}

\title{Chiral state conversion in a levitated micromechanical oscillator with \textit{in situ} control of parameter loops}

\author{Peiran Yin }
\affiliation{Hefei National Laboratory for Physical Sciences at the Microscale and Department of Modern Physics, University of Science and Technology of China, Hefei 230026, China}
\affiliation{CAS Key Laboratory of Microscale Magnetic Resonance, University of Science and Technology of China, Hefei 230026, China}
\affiliation{Synergetic Innovation Center of Quantum Information and Quantum Physics, University of Science and Technology of China, Hefei 230026, China}

\author{Xiaohui Luo  }
\affiliation{National Laboratory of Solid State Microstructures and Department of Physics, Nanjing University, Nanjing, 210093, China}

\author{Liang Zhang  }
\author{Shaochun Lin }
\author{Tian Tian  }
\affiliation{Hefei National Laboratory for Physical Sciences at the Microscale and Department of Modern Physics, University of Science and Technology of China, Hefei 230026, China}
\affiliation{CAS Key Laboratory of Microscale Magnetic Resonance, University of Science and Technology of China, Hefei 230026, China}
\affiliation{Synergetic Innovation Center of Quantum Information and Quantum Physics, University of Science and Technology of China, Hefei 230026, China}
\author{Rui Li }
\author{Zizhe Wang }
\affiliation{Hefei National Laboratory for Physical Sciences at the Microscale and Department of Modern Physics, University of Science and Technology of China, Hefei 230026, China}
\affiliation{CAS Key Laboratory of Microscale Magnetic Resonance, University of Science and Technology of China, Hefei 230026, China}
\affiliation{Synergetic Innovation Center of Quantum Information and Quantum Physics, University of Science and Technology of China, Hefei 230026, China}

\author{Changkui Duan }
\affiliation{Hefei National Laboratory for Physical Sciences at the Microscale and Department of Modern Physics, University of Science and Technology of China, Hefei 230026, China}
\affiliation{CAS Key Laboratory of Microscale Magnetic Resonance, University of Science and Technology of China, Hefei 230026, China}
\affiliation{Synergetic Innovation Center of Quantum Information and Quantum Physics, University of Science and Technology of China, Hefei 230026, China}
\author{Pu Huang }
\thanks{Corresponding author: \href{mailto:hp@nju.edu.cn}{hp@nju.edu.cn}}
\affiliation{National Laboratory of Solid State Microstructures and Department of Physics, Nanjing University, Nanjing, 210093, China}
\author{Jiangfeng Du }
\thanks{Corresponding author: \href{mailto:djf@ustc.edu.cn}{djf@ustc.edu.cn}}
\affiliation{Hefei National Laboratory for Physical Sciences at the Microscale and Department of Modern Physics, University of Science and Technology of China, Hefei 230026, China}
\affiliation{CAS Key Laboratory of Microscale Magnetic Resonance, University of Science and Technology of China, Hefei 230026, China}
\affiliation{Synergetic Innovation Center of Quantum Information and Quantum Physics, University of Science and Technology of China, Hefei 230026, China}

\begin{abstract}

Physical systems with gain and loss can be described by a non-Hermitian Hamiltonian, which is degenerated at the exceptional points (EPs). Many new and unexpected features have been explored in the non-Hermitian systems with a great deal of recent interest. One of the most fascinating features is that, chiral state conversion appears when one EP is encircled dynamically. Here, we propose an easy-controllable levitated microparticle system that carries a pair of EPs and realize slow evolution of the Hamiltonian along loops in the parameter plane. Utilizing the controllable rotation angle, gain and loss coefficients, we can control the structure, size and location of the loops \textit{in situ}. We demonstrate that, under the joint action of topological structure of energy surfaces and nonadiabatic transitions (NATs), the chiral behavior emerges both along a loop encircling an EP and even along a straight path away from the EP. This work broadens the range of parameter space for the chiral state conversion, and proposes a useful platform to explore the interesting properties of exceptional points physics.

\end{abstract}
\maketitle

\section{ INTRODUCTION}
Physical systems with non-conservative energy, which should be described by a non-Hermitian Hamiltonian, have attracted considerable research attentions in recent years \cite{el-ganainy2018non, miri2019exceptional, zdemir2019parity}. The gain and loss in these systems can cause the resonant modes to grow or decay exponentially with time. As a result, the norm of a wave function is no longer conserved and the eigenvectors are not orthogonal \cite{bender1998real, bender2007making, rotter2009a, heiss2012the}. For non-Hermitian Hamiltonian, the eigenvalue is extended into the complex field. Then the EP, where both eigenvalues and eigenvectors coalesce, can emerge on the intersecting Riemann sheets \cite{heiss1991avoided, heiss2000repulsion}. In the last few years, various counterintuitive features and fascinating applications have been explored in EPs physics, including loss-induced transparency \cite{jing2015optomechanically}, unidirectional invisibility \cite{lin2011unidirectional, feng2013experimental}, single-mode lasing \cite{hodaei2014parity, feng2014single}, band merging \cite{zhen2015spawning} and enhanced sensing \cite{wiersig2014enhancing, hodaei2017enhanced, chen2017exceptional, zhang2019quantum}. 

In particular, one of the most intriguing features of EP is that, after an adiabatic Hamiltonian evolution along a closed loop in the parameter space (called an adiabatic encirclement), the system does not return to its initial state, but to a different state on another Riemann sheet \cite{dembowski2001experimental, mailybaev2005geometric, lee2009observation, gao2015observation}. And after a second encirclement, it would return to the initial state. Later, it was found that the adiabatic prediction breaks down in such an encirclement, because the system is singularly perturbed by the nonadiabatic couplings \cite{uzdin2011on, berry2011slow, gilary2013time, milburn2015general}. In a fully dynamical picture, additional nonadiabatic transitions together with the topological structure of EP give rise to the fascinating chiral behavior that the direction of encirclement alone determines the final state of the system. This dynamical encircling process has initiated intense research efforts and has been studied both theoretically in different frameworks \cite{hassan2017dynamically, wang2018non} and experimentally in microwave arrangements \cite{doppler2016dynamically}, optomechanical system \cite{xu2016topological} and coupled optical waveguides \cite{yoon2018time}.

Recently, it has been found that the chiral state conversion is conditional. The start point of the loop can affect the dynamics of chiral state conversion \cite{zhang2018dynamically, liu2020dynamically}, i.e., start point in the broken $PT$ symmetric phase leads to nonchiral dynamics. And even along a loop excluding the EP but in the vicinity of EP, chiral behavior appears \cite{hassan2017chiral, hassan2017erratum}. Besides, the homotopic loops with different shapes can also lead to distinct outcomes \cite{zhang2019distinct}. The NAT in the dynamical process that is of fundamental interest is the key to the chiral dynamics. However, the influence of structure, size and location of the parameter loop on the dynamics is still lack of exploration, especially experimentally.

 Here, we propose a new platform to realize the controllable dynamical evolution loop in a levitated micromechanical oscillator. In this platform, we make the recently discussed dynamical features of EPs directly accessible through \textit{in situ} control of the system parameters, i.e., the rotation angle, gain and loss coefficients. We study the influence of structure, size and location of the parameter loop on the chiral state conversion process, so the range of parameter space for the experimental realization of which is broadened.

 \section{ LEVITATED MICROPARICLE SYSTEM }

 A diamagnetic microparticle can be levitated stably under a balance of diamagnetism force and gravity, which has already been realized with superconductor \cite{geim1999magnet} or permanent magnets \cite{slezak2018cooling, PhysRevResearch.2.013057}. As shown in Fig.~1(a), the magneto-gravitational trap generated by a set of permanent magnets can be described as a harmonic potential near the equilibrium position, i.e., $U(\bm{r})= \frac{1}{2}k_xx^2+\frac{1}{2}k_yy^2+\frac{1}{2}k_zz^2$, where $k_x$, $k_y$ and $k_z$ are the spring constants in the $x$, $y$ and $z$ directions, respectively. Thus the motion of a levitated microparticle can be regarded as three separated resonant modes with frequencies being $\omega^2_{x0}=k_x/m$, $\omega^2_{y0}=k_y/m$ and $\omega^2_{z0}=k_z/m$. Here, $m$ is the mass of the microparticle.

   \begin{figure}
	\centering
	\includegraphics[width=0.95\columnwidth]{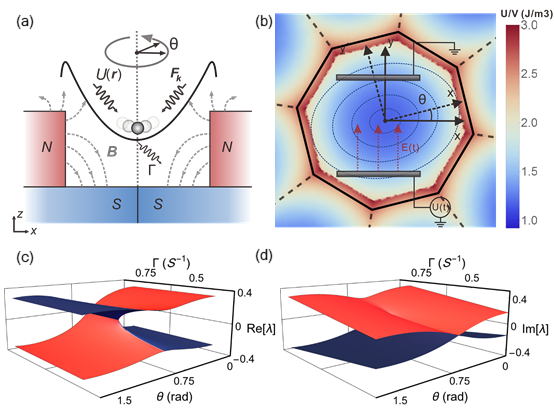}
	\caption{ 
		(a) Schematic plot of the levitated microparticle system. The diamagnetic microparticle in the magneto-gravitational trap can be described as a three-mode resonator near the equilibrium. The magnets set can be rotated around the $z$ axis. The gas damping of the motion in $x$ direction is $\Gamma$.
		(b) Schematic plot of the $x$ and $y$ motion modes. The bound field is rotated by an angle $\theta$,  with the potential energy density indicated by color. A pair of electrodes are used to conduct the feedback control of the motion mode in $y$ direction.
		(c), (d) Calculated real part (c) and imaginary part (d) of the eigenvalues as a function of the gain or loss $\Gamma$ and the rotation angle $\theta$. The self-intersecting Riemann surfaces represent the two branches of eigenvalues, with red referring to the gain eigenstate and blue referring to the loss eigenstate. And an EP is marked with an arrow.
	}\label{system}
\end{figure}

 Henceforth, we treat the microparticle motion in $x$ and $y$ directions as two distinct motion modes with resonance frequencies $\omega_{x0}$ and $\omega_{y0}$, respectively. Then we rotate the magneto-gravitational bound field around the $z$ axis by an angle $\theta$, as shown in Fig.~1(b). But we still select the $x$ and $y$ motion states as the modes. So the bound field can still be expressed in the $x$-$y$ coordinate as:
  \begin{equation}
 \begin{split}
 U(\bm{r})&=\frac{1}{2}k_x(x\cos\theta+y\sin\theta)^2\\
      &+\frac{1}{2}k_y(-x\sin\theta+y\cos\theta)^2.
 \end{split}
 \end{equation}
 As a result, the resonance frequencies of $x$ and $y$ modes become $ \omega_x^2=(\omega^2_{x0}+\omega^2_{y0})/2+\cos2\theta(\omega^2_{x0}-\omega^2_{y0})/2 , \omega_y^2=(\omega^2_{x0}+\omega^2_{y0})/2-\cos2\theta(\omega^2_{x0}-\omega^2_{y0})/2 $.
 
 The loss of energy in this system comes from gas damping. For the $x$ mode motion, the loss coefficient is indicated by $\Gamma_x$, which can be controlled by tuning the pressure. And for the motion in $y$ direction, we conduct the feedback control using a pair of electrodes, as shown in Fig.~1(b). We conduct a real-time measurement of the microparticle's motion with a CCD camera. Then we apply an electric force proportional to $\dot{y}$ on the microparticle in the $y$ mode. Such a feedback control of the $y$ mode effects as a gain coefficient $\Gamma_y$. 
 
 As a result, the dynamical equation for the two motion states is:
 \begin{equation}
 \begin{split}
 \ddot{x}+\Gamma_x\dot{x}+\omega_x^2x+\eta{y}&=0\\
 \ddot{y}-\Gamma_y\dot{y}+\omega_y^2x+\eta{x}&=0,
 \end{split}
 \end{equation} 
 with $\eta=\frac{1}{2}(\omega^2_{x0}-\omega^2_{y0})\sin2\theta$. We define the average uncoupled-resonance angular frequency to be $\omega_0=(\omega_{x0}+\omega_{y0})/2$, and the detuning to be $\Omega=\omega_{x0}-\omega_{y0}$. Here, two issues need to be addressed. First, different from typical mode-coupling system with a Hermitian Hamiltonian, the two motion modes in this model are in a new interaction scheme: $\Gamma_x, \Gamma_y\approx\eta/\omega_0\approx\Omega\ll\omega_{x0}, \omega_{y0}$. Hence, we can get the coupling peaks as we drive and measure the motion modes in $x$ and $y$ direction, without a requirement for parametric modulation. Second, we conduct the feedback control of the motion in $y$ direction, reflecting our selection of the motion mode in a physical sense.

 In the weak-coupling and small-detuning regimes satisfying $\eta\ll\omega_n$ and $\Omega\ll\omega_0$, we can simplify Eq.~(2) with a slowly-varying complex-envelope function $A_n(t)$, such that, $x(t)=A_x(t)e^{\mathrm{i}\omega_0t}+A_x^*e^{-\mathrm{i}\omega_0t}$ and $y(t)=A_y(t)e^{\mathrm{i}\omega_0t}+A_y^*e^{-\mathrm{i}\omega_0t}$. Then, using a vector $\bm{\Psi}=(A_x,A_y)^T$ to describe $x$ and $y$ motion state of the system, we obtain a Schr\"odinger-type coupled-mode equation: 
 \begin{align} 
\mathrm{i}\partial_t{\bm{\Psi}}=H\bm{\Psi}.
\end{align}
 And the effective two-state Hamiltonian is given by
 \begin{align} 
  	H={\left(\begin{array}{cccc} 
  	-\mathrm{i}\Gamma_x/2-\frac{\Omega}{2}\cos2\theta&-\frac{\Omega}{2}\sin2\theta\\ -\frac{\Omega}{2}\sin2\theta&\mathrm{i}\Gamma_y/2+\frac{\Omega}{2}\cos2\theta)
  	\end{array}\right)}.
\end{align} 
Since it is always possible to remove the trace of $H$ by a simple gauge transformation \cite{guo2009observation}. Here, we simplify $H$ by making $\Gamma_x\approx \Gamma_y=\Gamma$, without loss of generality. And we define the effective detuning and coupling as $\Omega_{\textrm{eff}}=\frac{\Omega}{2}\cos2\theta$ and $\eta_{\textrm{eff}}=\frac{\Omega}{2}\sin2\theta$, respectively.

For the two-mode system, the time-reversal operation $T$ transforms a time-independent operator to its complex conjugate, while the parity operator $P$ exchanges locations of the modes. Thus, it is easy to verify that our model Hamiltonian becomes $PT$-symmetric when the effective detuning $\Omega_{\textrm{eff}}=\frac{\Omega}{2}\cos2\theta$ is zero. The eigenvalue of this Hamiltonian is $\lambda_\pm=\pm\frac{1}{2}\sqrt{\Omega^2-\Gamma^2+2\mathrm{i}\Omega\Gamma\cos2\theta}$. The EPs emerge when $\Omega=\Gamma$ and $\cos2\theta=0$, that is, a pair of EPs locate at $\Gamma=\Omega$ and $\theta=\pi/4, 3\pi/4$, respectively. Figs.~1(c) and 1(d) show the calculated real part and imaginary part of the eigenvalues over the parameter plane of $\Gamma$ and $\theta$ in the vicinity of the EP at $\theta=\pi/4$. We can find that, $\lambda_-$ and $\lambda_+$ coalesce at the EP, and in the vicinity of the EP, they exhibit the same structure as Riemann sheets of complex square-root function $z^{\frac{1}{2}}$. The $PT$ symmetric phase line is a branch cut that connects the gain Riemann sheet (see the red sheet in Fig. 1(c)) with the loss Riemann sheet (see the blue sheet in Fig. 1(c)). 
  \begin{figure}
	\centering
	\includegraphics[width=0.95\columnwidth]{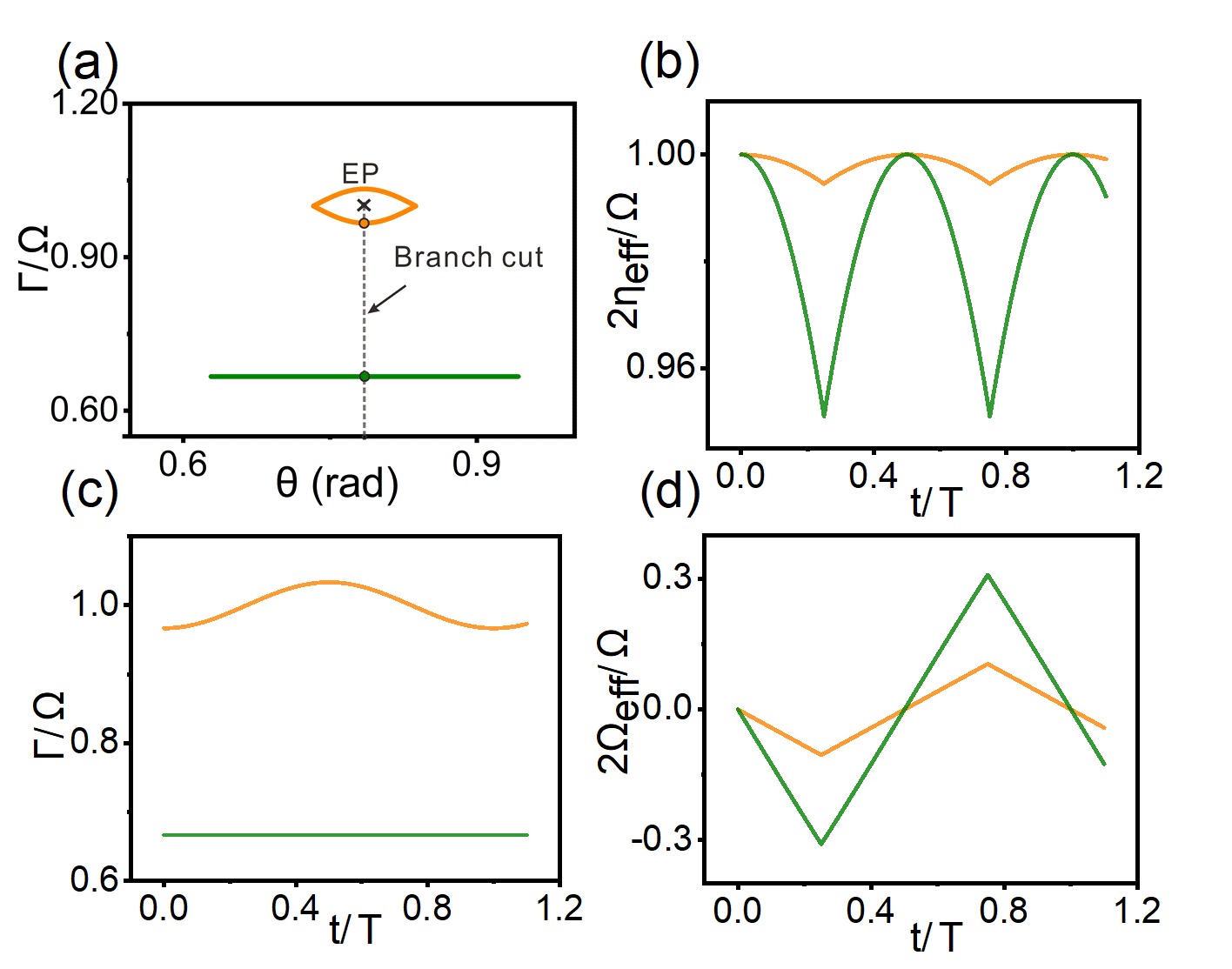}
	\caption{ 
		(a) The orange loop in the parameter plane of $\theta$ and $\Gamma$ is encircling an EP at $\theta=\pi/4$ and $\Gamma=\Omega$, marked by a cross. The green line indicates a straight path away from the EP with $\Gamma_0=2\Omega/3$. The start points indicated by the colored dots are both on the branch cut, which is the $PT$ symmetric phase. The parameters we have chosen are accessible in the levitated  micromechanical oscillator, with $\Omega=2\pi\times0.1 (S^{-1})$, $\omega_c=\Omega/8$ and $A_\Gamma = \Omega/30$.
		(b)-(d) The CCW evolution of the time-dependent parameters $\eta_{\textrm{eff}}$, $\Gamma$ and $\Omega_{\textrm{eff}}$ in one period $T$ along the orange and green loops.
	}\label{param}
\end{figure}

 In this magnetogravitational trap system, microparticle with diameter between $2\mu$m and 1mm can be trapped stably at room temperature. The typical parameters $\Omega\approx 2\pi\times0.1 \textrm{s}^{-1}$, $\omega_0\approx2\pi\times 10 \textrm{s}^{-1}$ is accessible in the system. To approach the EP, we have to make the gain or loss coefficient $\Gamma$ around $2\pi\times0.1 \textrm{s}^{-1}$, which is achievable at a pressure of $P\sim10^{-3}$ mbar for a microparticle with diameter around $10\mu$m. In such a low-frequency system, stable trapping and effective motion detection have been well realized \cite{slezak2018cooling, PhysRevResearch.2.013057}. We can use a 633-nm laser as the illumination source and use an objective to collect scattered light from the microsphere. The position of microparticle is tracked with a CCD camera with a speed of 200 frames per second. 
 
 To realize the dynamical evolution of Hamiltonian $H$ in the parameter space, the gain or loss coefficient $\Gamma$ and the rotation angle $\theta$ have to be controlled \textit{in situ} coherently with high precision. We can modulate the rotation angle $\theta$ uniformly and slowly with a constant angular frequency. Then the parameter $\theta$ oscillates between $[-A_\theta, A_\theta]$, with a constant rate $\frac{2A_\theta}{\pi}\omega_c$.
\begin{align}
&\theta=\frac{\pi}{4}+\frac{2A_\theta}{\pi}\int_0^t f(t')\omega_c \mathrm{d} t' \textrm{, with}\\
&f(t')=\left\{ \begin{array}{ll}
1 & \textrm{if $\cos\omega_c t'\geq 0$}\\
-1 & \textrm{if $\cos\omega_c t'<0.$} \notag
\end{array} \right. 
  \end{align} 
  Due to the slow evolution of our system, during the rotation, we can also coherently control the feedback amplification and gas damping of the particle:
  \begin{align}
 \Gamma=\Gamma_0+A_\Gamma\sin(\omega_c t+\pi/2).
  \end{align}
In this experimental scheme, $\Gamma$ is in the range of $2\pi \times0.01 \textrm{s}^{-1}$ to $2\pi \times 0.1 \textrm{s}^{-1}$, which is much larger than the frequency fluctuations due to the changes of external environmental conditions. Thus we can effectively control the gain coefficient.
  
  In the experimental system, considering the effective range of the stable trapping, $\omega_c\succeq \Omega/15$ is achievable. In principle, $0\leq A_\theta\leq \pi/2$ and $2\pi\times10^{-4} \textrm{s}^{-1}\leq A_\Gamma\leq \Gamma_0$ is achievable in the system, and it is a range of parameters sufficient to realize the parametric evolution of the Hamiltonian. Hence, parameter loops with different structure, size and location are accessible in our system. As shown in Fig.~2(a), in the parameter plane of $\Gamma$ and $\theta$, the orange loop is encircling an EP with $A_\theta= \pi/30$ and $A_\Gamma=\Omega/30$, while the green loop has the simplified structure being a straight path away from the EP with $A_\theta= \pi/10$ and $A_\Gamma=0$. The start points are both at the branch cut. The corresponding slow evolution of parameters $\eta_{\textrm{eff}}$, $\Gamma$ and $\Omega_{\textrm{eff}}$ along the loops in one period $T=2\pi/\omega_c$ is shown in Figs.~2(b)-2(d). We define that the parameter loop is counterclockwise (CCW) when $\omega_c>0$, and clockwise (CW) when $\omega_c<0$. In this dynamical evolution process, our system serves as an easy-controllable platform with $\Gamma_0$, $A_\Gamma$, $\omega_c$ and $A_\theta$ all being tunable. So the size and location of the parameter loop can be coherently controlled \textit{in situ} to study their effects on the chiral state conversion process.

 \section{ CHIRAL STATE CONVERSION IN DYNAMICAL ENCIRCLEMENT }
 
 The dynamics of the system is now definitely determined by Eq.~(3), with the non-Hermitian Hamiltonian being time-dependent. The instantaneous eigenbasis of such a Hamiltonian is not orthogonal in the sense of Dirac. Instead, a biorthogonal eigenbasis can be constructed with right eigenvectors $ \bm{r_+} $, $\bm{r_-}$ and corresponding left eigenvectors $\bm{l_+}$, $\bm{l_-}$, where the subscripts $+$ and $-$ denote the eigenstates with gain and loss respectively. They are defined by $H\bm{r_\pm}=\lambda_\pm\bm{r_\pm}$ and $\bm{l_\pm}^TH=\lambda_\pm\bm{l_\pm}^T$, such that $\bm{l_i}^T\bm{r_j}=\delta_{i,j}.$ To further study the dynamical process, we choose the parallel transported eigenbasis \cite{milburn2015general}:
 \begin{align*} 
 \bm{r_+}=\bm{l_+}={\left(\begin{array}{cccc} 
 	-\sin(\alpha/2)\\\cos(\alpha/2) 
 	\end{array}\right)},
  \bm{r_-}=\bm{l_-}={\left(\begin{array}{cccc} 
 	\cos(\alpha/2)\\\sin(\alpha/2)
 	\end{array}\right)},
 \end{align*}
 with $\alpha$ defined by $\tan\alpha=\frac{\Omega}{2}\sin2\theta/(\frac{\Omega}{2}\cos2\theta+\mathrm{i}\Gamma/2)$. Then, any vector state at time $t$, can be expanded into a linear combination of the eigenbasis (i.e., right eigenvectors): 
\begin{align}
\bm{\Psi}=c_+\bm{r_+}+c_-\bm{r_-}.
\end{align}

In all cases we set the start point of the parameter loop at the branch cut. And without loss of generality, we let the initial state be one of the eigenstates $\bm{r_\pm}$. Then we study the dynamical evolution of coefficients $(c_+, c_-)$, whose initial value is $(1,0)$ or $(0,1)$ in the instantaneous basis. In the numerical studies shown in the following, we have seen indications consistent with previous theoretical result that adiabatic behavior for at least one state is not always observed \cite{nenciu1992on}.

Firstly, we study the case that the parameter loop encircles an EP. As shown in Fig.~3, the dynamical encirclement around an EP with initial state prepared to $\bm{r_{+}}(0)$ or $\bm{r_-}(0)$ is calculated numerically. When the encircling direction is CW (indicated by the arrow on the trajectory), both $\bm{r_+}(0)$ and $\bm{r_-}(0)$ will evolve to $\bm{r_-}(0)$ after one encirclement, as shown in Figs.~3(a) and 3(b), respectively. In Fig.~3(a), this evolution matches the adiabatic prediction showing a state-flip, which is a direct result from the topological structure around EP. While in Fig.~3(b), a NAT appears and leads to a sudden state switch \cite{gong2019piecewise}. That is, the adiabatic following dynamics is piecewise. With the combined action of the topological structure around EP (the adiabatic evolution around EP) and the appearance of a NAT (a sudden state switch), the final state goes back to the initial state $\bm{r_-}(0)$. Here, we define the condition $\vert c_-\vert=\vert c_+\vert$ as the confirmation of appearance of a NAT. And in Figs.~3(c) and 3(d), the encircling direction is CCW, both $\bm{r_+}(0)$ and $\bm{r_-}(0)$ evolve to $\bm{r_+}(0)$ after one encirclement. To summarize, the final state of such a dynamical encirclement depends on the evolution direction, this asymmetric mode switch is called chiral state conversion.
 \begin{figure}
	\centering
	\includegraphics[width=0.95\columnwidth]{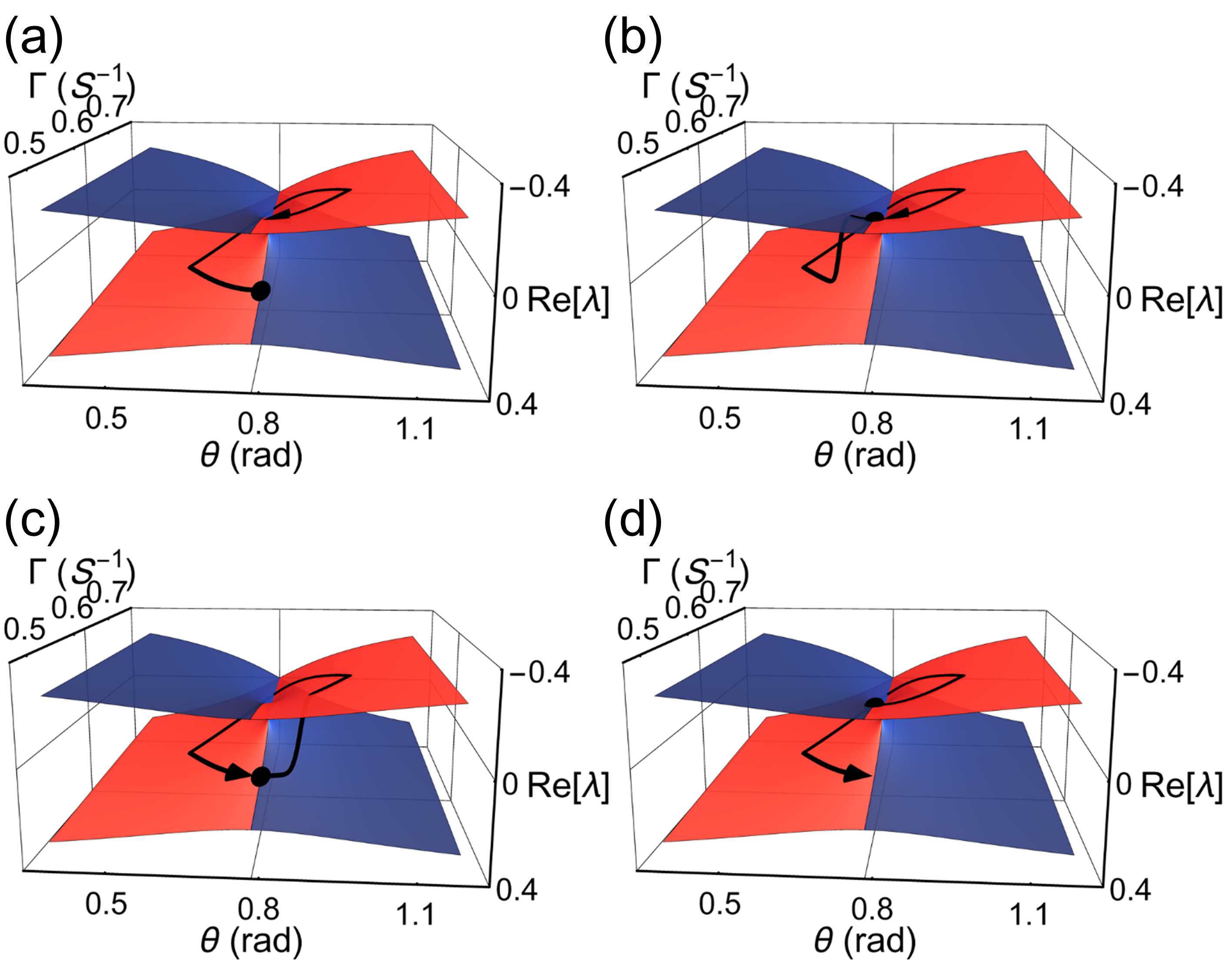}
	\caption{ 
		(a,~b) The CW encircling trajectories on the intersecting Riemann sheets. The initial states are $\bm{r_+}(0)$ in (a) and $\bm{r_-}(0)$ in (b), respectively. The start point is indicated by a dot and the trajectory is calculated by $(\vert c_-(t)\vert^2\lambda_-(t)+\vert c_+(t)\vert^2\lambda_+(t))/(\vert c_-(t)\vert^2+\vert c_+(t)\vert^2)$.
		(c,~d) The CCW encircling trajectories on the intersecting Riemann sheets.
	}\label{loopline}
\end{figure}

To further understand the physics in this dynamical process, we perform numerical simulations to study the nonadiabatic transitions in the slow encirclement. With encircling direction being CW (the cases shown in Figs.~3(a) and 3(b)), the amplitudes of $c_+$ and $c_-$ are shown in Figs.~4(a) and 4(b). In Fig.~4(a), the initial state is $\bm{r_+}(0)$. Since the state always evolves on the gain Riemann sheet (see also Fig.~3(a)), it is stable and $c_+$ dominates in the whole process. In Fig.~4(b), the initial state is $\bm{r_-}(0)$, so the state evolves on the loss Riemann sheet (see also Fig.~3(b)) at the beginning. But a NAT happens after $t_d$, which is the delay time counted from the last time (the beginning in this case) that the loop goes across the branch cut into the loss state \cite{milburn2015general}, thus the state switches to the gain state $\bm{r_+}(t)$. This transition can be understood  physically as that, if not being the lowest loss one, a state of the non-Hermitian system evolving in a parameter loop is unstable, and would transfer to the lowest loss state as long as the evolution time is sufficiently long. As a result, the final state is always $\bm{r_-}(0)$ in the CW direction, no matter what the initial state is.     
\begin{figure}
	\centering
	\includegraphics[width=0.95\columnwidth]{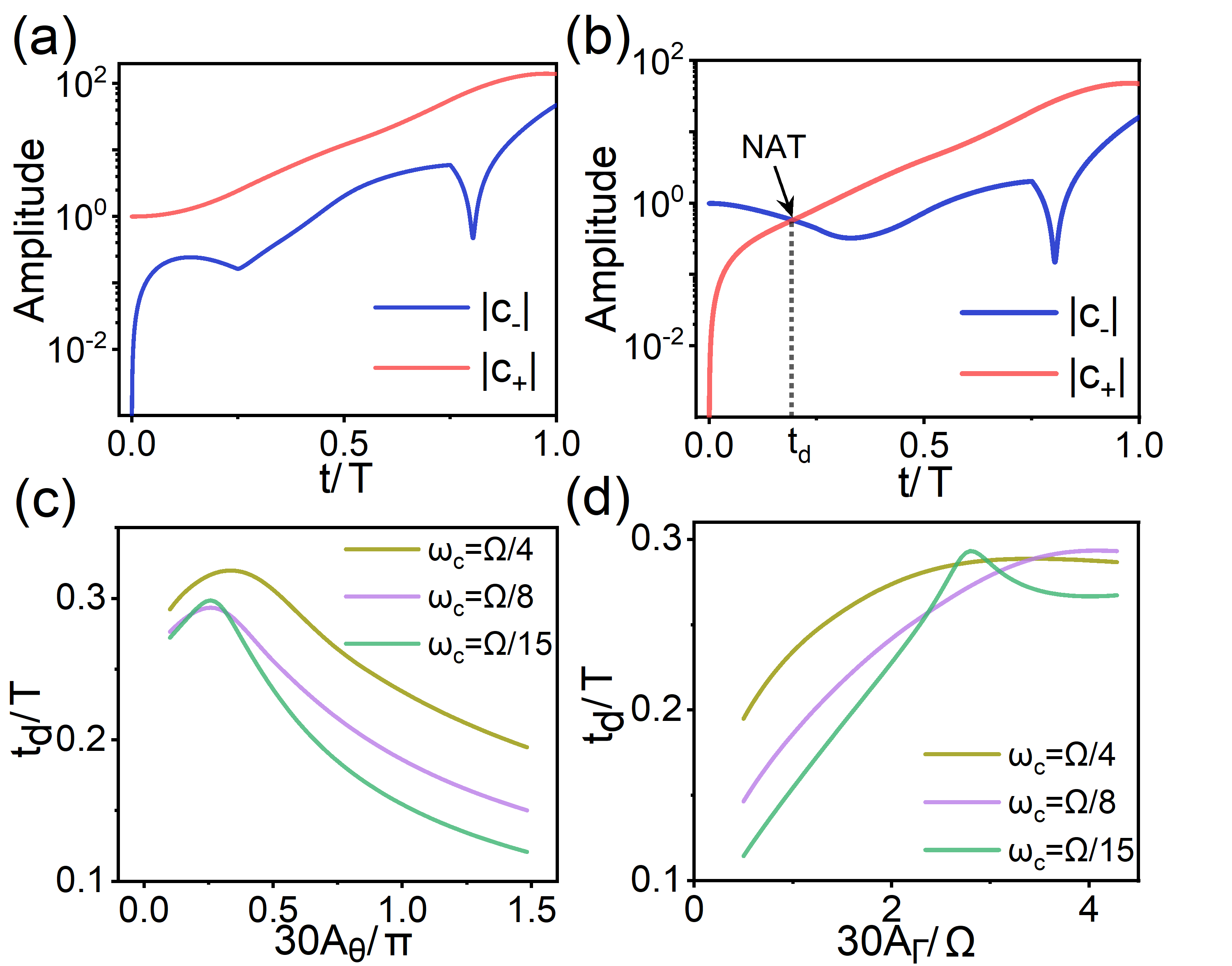}
	\caption{ 
		(a,~b) Calculated amplitudes of $c_+$ and $c_-$ in the CW evolution, with initial state being $\bm{r_+}(0)$ in (a) and $\bm{r_-}(0)$ in (b). We choose the parameters to be $\omega_c=\Omega/8$, $A_\Gamma=\Omega/30$ and $A_\theta=\pi/30$.
		(c) The delay time $t_d$ as a function of $A_\theta$, with different $\omega_c$'s and $A_\Gamma=\Omega/30$.
		(d) The delay time $t_d$ as a function of $A_\Gamma$, with different $\omega_c$'s and $A_\theta=\pi/30$.
	}\label{tdelay}
\end{figure}

In our system, $A_\theta$, $A_\Gamma$ and $\omega_c$ are all tunable, so we can study their influence on $t_d$. The numerical results are given in Figs.~4(c) and 4(d). We find that $t_d$ is always smaller than $T/2$. In such an encirclement, we can realize the chiral state conversion if $t_d<T$, which is met robustly in our system. Although these parameters determining the loop size can affect $t_d$ slightly, the chiral state conversion is achieved robustly.

\section{ CHIRAL STATE CONVERSION  ALONG A STRAIGHT PATH IN PARAMETER SPACE }
 
 Secondly, we simplify the structure of parameter loop to be a straight path across the branch cut by setting $A_\Gamma=0$ (see also Fig.~2(a)). The start point of the loop is still at the brunch cut ($\theta$=$\pi$/4), but now $\Gamma_0$ is variable. Our numerical calculation shows that, without the limit of small loop scale, we can achieve chiral state conversion by increasing length of the straight path $2A_\theta$, which has been preliminarily explored in Fig.~4(c). The four dynamical evolution processes are shown in Fig.~5. We can see that, even the path is away from the EP with $\Gamma_0=\frac{2}{3}\Omega$, the final state of one loop is always $\bm{r_-}(0)$ in CW direction (see Figs.~5(a) and 5(b)) and $\bm{r_+}(0)$ in CCW direction (see Figs.~5(c) and 5(d)). These results are in accordance with those along a loop encircling an EP (see Fig.~3), that is, the same chiral behavior. 
 
 \begin{figure}
 	\centering
 	\includegraphics[width=0.95\columnwidth]{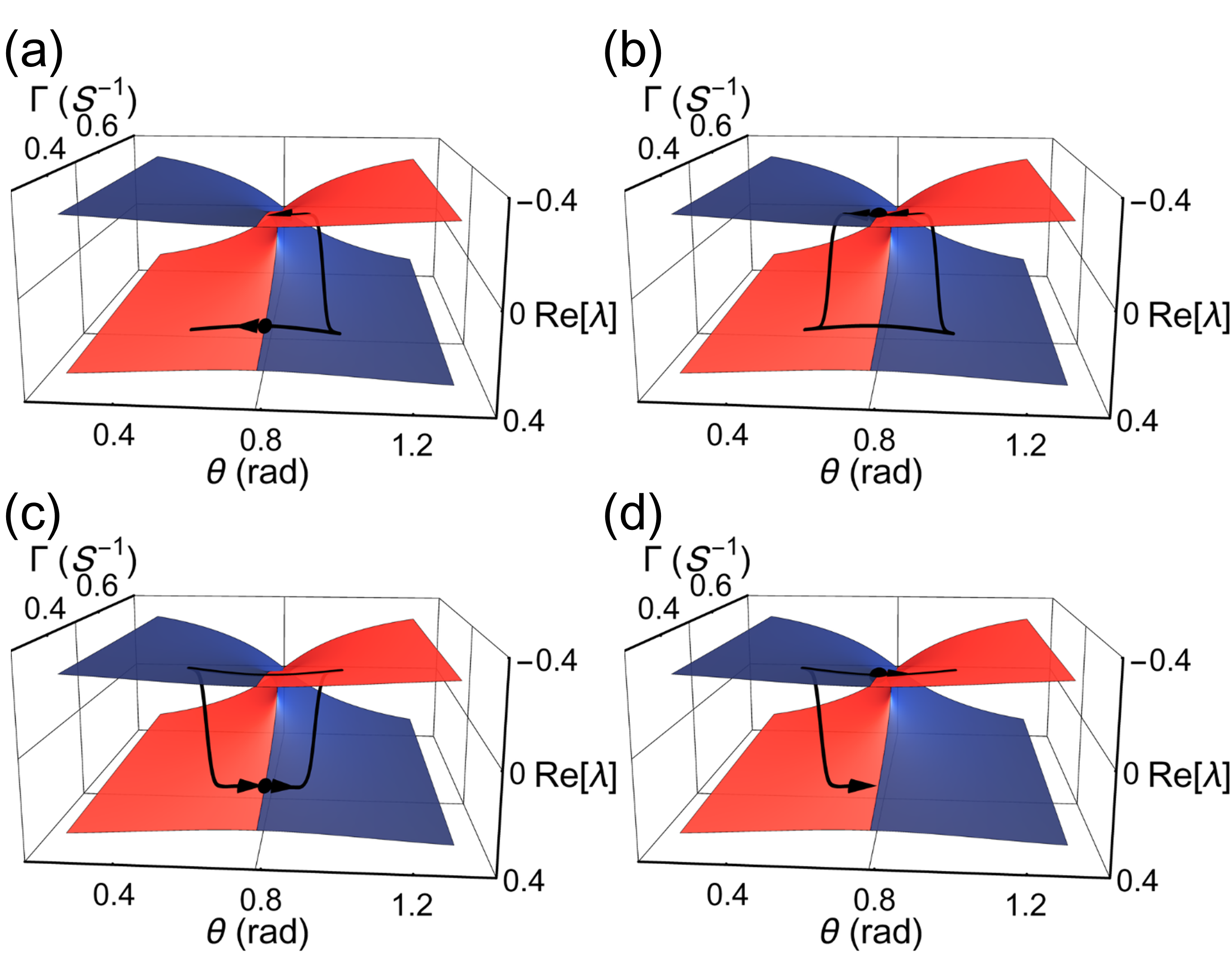}
 	\caption{ 
 		(a,~b) The CW trajectories on the intersecting Riemann sheets for the straight path in parameter plane. The initial states are $\bm{r_+}(0)$ in (a) and $\bm{r_-}(0)$ in (b), respectively. The start point is indicated by a dot, near which is an arrow to show the loop direction at the beginning.
 		(c,~d) The CCW trajectories on the intersecting Riemann sheets.
 	}\label{loopline1}
 \end{figure}

 In order to analyze the evolution processes in Figs.~5(a) and 5(b), the numerical results of the time-dependent coefficients $\vert c_+\vert$ and $\vert c_-\vert$ in the CW dynamical processes are plotted in Figs.~6(a) and 6(b). We find that the straight path goes across the branch cut at $T/2$, and after crossing it, the two states have switched with each other.  As shown in Fig.~6(a), the state evolves adiabatically in the first half of the evolution process, while after the branch cut a NAT happens with $t_d(a)$ due to the instability of loss state in the evolution. Therefore the final state switches to $\bm{r_-}(0)$, which is the same result as that shown in Fig.~4(a), but has different dynamics. And in Fig.~6(b), two NATs emerge with corresponding delay times $t_d(b1)$ and $t_d(b2)$ because the initial state is the loss state $\bm{r_-}(0)$, which is unstable from the beginning.   \begin{figure}
 	\centering
 	\includegraphics[width=0.95\columnwidth]{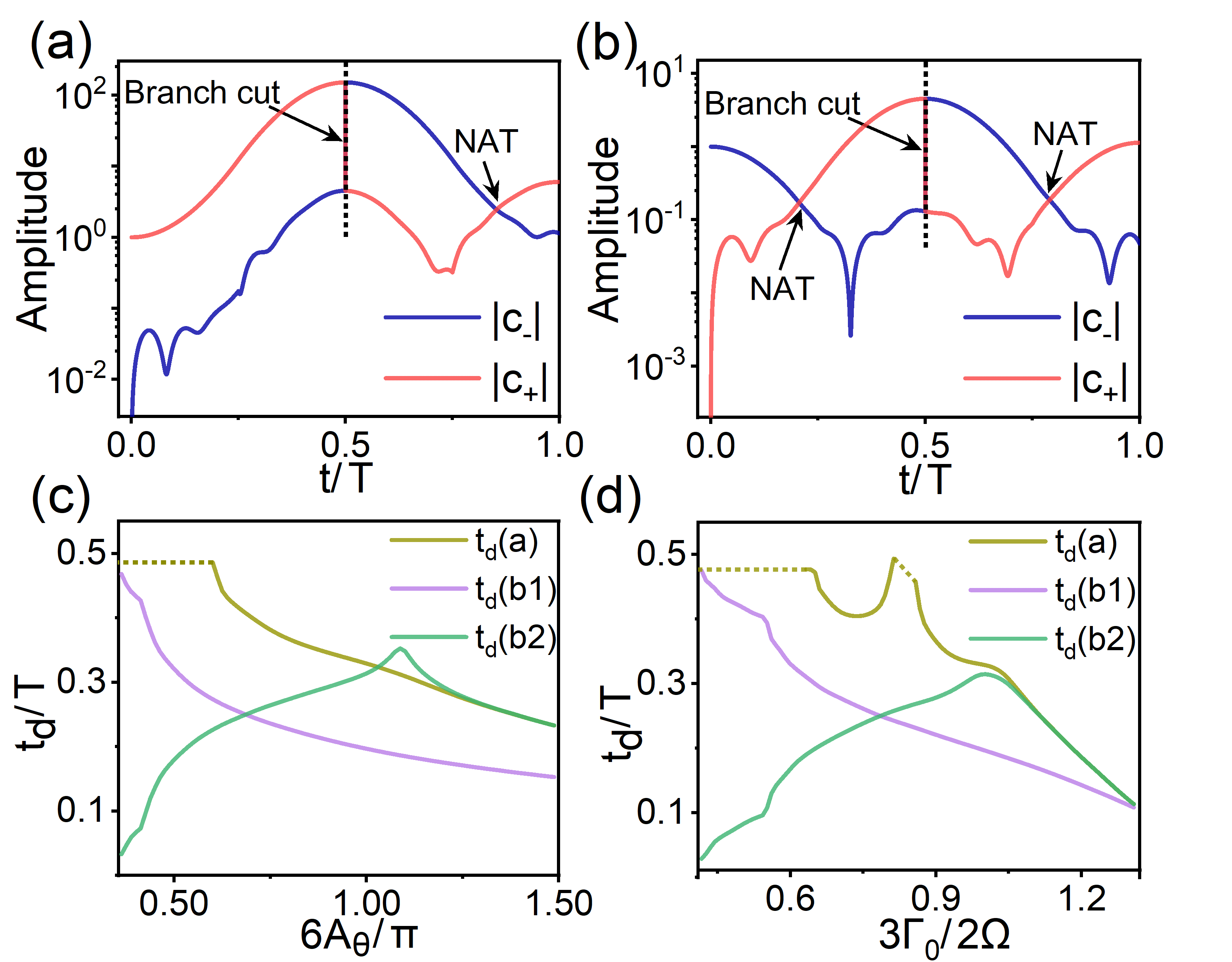}
 	\caption{ 
 		(a,~b) Calculated amplitudes of $c_+$ and $c_-$ in the CW evolution, with initial state being $\bm{r_+}(0)$ in (a) and $\bm{r_-}(0)$ in (b). We choose the parameters to be $\omega_c=\Omega/17$, $A_\theta=\pi/6$ and $\Gamma_0=2\Omega/3$. There is a branch cut at $T/2$ in each panel as marked.
 		(c) The delay times $t_d(a)$, $t_d(b1)$ and $t_d(b2)$ as a function of $A_\theta$ with $\Gamma_0=2\Omega/3$.
 		(d) The delay times $t_d(a)$, $t_d(b1)$ and $t_d(b2)$ as a function of $\Gamma_0$ with $A_\theta=\pi/6$.
 	}\label{tdelayline1}
 \end{figure}
After the first NAT the state jumps to the stable gain state, while after the branch cut it switches to the unstable loss state again, which gives rise to the second NAT.  As a result, the final state switches two times back to the initial state $\bm{r_-}(0)$: also the same result as that in Fig.~4(b) but different dynamics. Hence, the chiral state conversion is also achieved in the straight path evolution but with different dynamics.

 Furthermore, we study the influence of parameters $A_\theta$ and $\Gamma_0$ on the delay times $t_{d}(a)$, $t_{d}(b1)$ and $t_{d}(b2)$ as shown in Figs.~6(c) and 6(d). The dashed line in the panel indicates the interval that no NAT happens. Under the condition of $t_d<T/2$, the chiral state conversion would appear. As shown in Fig.~6(c), we find that, increasing the range of the straight path $2A_\theta$ can make the chiral state conversion more robust. And we can see in Fig.~6(d) that, when the loop is approaching the EP ($\Gamma_0$ approaching $\Omega$), the chiral behavior becomes very robust as $t_d$ is becoming very small. And when the path is away from the EP, $t_d$ is not always smaller than $T/2$. That is, such chiral behavior is not so robust as that in the vicinity of an EP, but we can still find the accessible parameter space to achieve it (by increasing $A_\theta$).

\section{ CONCLUSION }
 
In summary, we have proposed an easy-controllable levitated micromechanical oscillator as a platform to study the slow evolution dynamics in different parameter loops. With \textit{in situ} control of the parameters $A_\theta$, $A_\Gamma$ and $\Gamma_0$, we have realized the chiral state conversion, both along a loop encircling an EP and a straight path away from the EP, with different dynamics in the process. It is a combination of the topological structure of energy surfaces and the NAT that leads to the chiral behaviors. We have broadened the range of parameter space for the chiral state conversion, that with much lower loss coefficient $\Gamma$, chiral state conversion is also realized with different dynamical process. Furthermore, we can use this platform to study the complicated dynamical processes governed by time-dependent non-Hermitian Hamiltonians, such as, the encircling of high-order EPs \cite{demange2012signatures, zhang2019dynamically}, non-Hermitian topological invariants \cite{longhi2019topological, okuma2020topological, xiao2020non} and floquet non-Hermitian physics \cite{yang2019floquet, li2019observation}.

\begin{acknowledgements}	
This work was supported by the Fundamental Research Funds for the Central Universities (Grant No. WK2030000032), National Key R$\&$D Program of China (Grant No. 2018YFA0306600), the CAS (Grant Nos. GJJSTD20170001 and QYZDY-SSW-SLH004), and Anhui Initiative in Quantum Information Technologies (Grant No. AHY050000).	
\end{acknowledgements}

\bibliography{chiralstateconversion_submission}

\end{document}